\documentclass[halfparskip]{scrartcl}

\usepackage[utf8]{inputenc}
\usepackage{amsmath}
\usepackage{amsfonts}
\usepackage{mathtools}
\usepackage{algorithm2e}
\usepackage{varioref}
\usepackage{tikz}
\usepackage{wrapfig}
\usepackage[numbers]{natbib}

\usepackage{hyperref}

\usetikzlibrary{shapes.geometric}
\usetikzlibrary{calc}

\author{Joachim Breitner\footnote{e-mail: \href{mailto:mail@joachim-breitner.de}{mail@joachim-breitner.de}}}
\title{Conditional Elimination through Code Duplication}
\begin{document}
\maketitle

\begin{abstract}
We propose an optimizing transformation which reduces program runtime at the expense of program size by eliminating conditional jumps.
\end{abstract}


\section{Preface}

\subsection{Motivation}

In a variety of cases, code is written in a way that in one execution, a conditional execution is evaluated several time. Situations where this may be happening include the following:

\begin{itemize}
\item Repeated use of the ternary operator \texttt{($\cdot$?$\cdot$:$\cdot$)} with a common conditional expression.
\item An if-then-else statement inside a loop, where the condition is loop invariant.
\item Use of macros or inlined functions provided by a library that include conditional expression.
\item Conditional jumps implicitly inserted by the compiler due to short-circuit logic.
\item Naive code mechanically generated from another source via tools such as parser generators, or compilers of higher languages that compile to C and then invoke a C compiler.
\item Conditionals introduced by earlier compilation passes, such as the Partial Dead Code pass conceived by Bodík and Gupta \citep{PDE} are likely to make other conditionals redundant. In fact, the PDE paper recommends a “branch elimination” step without giving the details of this. CECD can serve as an implementation of this step.
\end{itemize}

In some of these cases the programmer might be able to eliminate the redundant conditional expression by himself, but often at the cost of less readable code or repetition, such as two instances of the loop mentioned in the second bullet. In other cases, such as the library-provided macros or the generated code, it is not feasible to expect the source code to be free of redundant conditionals. Therefore it is desirable that an optimizing compiler can perform this transformation.

Furthermore, this transformation not only reduces execution time but can enabled further optimizations: If the conditional expression is of the form $v==c$ for a variable $v$ and a constant $c$, a constant propagation pass can replace $v$ by $c$ in the then-branch, which has been enlarged by our optimization. Also, modern computer architectures, due to long pipelines, perform better if fewer conditional jumps occur in the code.

\subsection{Outline}

In the next section, we explain when a given region to duplicate is valid and how to perform the conditional elimination. Aiming for a very clear, simple and homogeneous presentation, we describe the algorithm in a very general setting. This will possibly introduce dead code. An implementation would either run a dead code elimination pass afterwards or refine the given algorithm as required. The transformation is demonstrated by example.

Section \ref{sec:region} discusses which properties the region should satisfy for the optimization to actually have a positive effect, and how to avoid useless code duplication.

To decide whether to perform the optimization, we give a simple heuristic that selects a region to be duplicated and decides whether the optimization should be performed, weighting the (runtime) benefits weighted against the (code size) cost in subsection \ref{sec:heurisitc}. We also show that a slight more sophisticated approach, which takes profiling information into account, becomes $\mathcal{NP}$-hard.

Data flow equations for the properties discussed in the preceding two sections are given in \ref{sec:df}.

\subsection{Acknowledgements}

This paper was written for the group project of the CS614 “Advanced Compiler” course at IIT Bombay under Prof. D. M. Dhamdhere. I have had fruitful discussions with him and my fellow group members, Anup Agarwal, Yogesh Bagul and Anup Naik, who subsequently implemented parts of this using the LLVM compiler suite.

\section{Conditional elimination}

Let $e$ be an expression, which should occur as the condition for a conditional branch in the control flow graph (CFG) of a program, and let $v_1,\,v_2,\,\ldots$ be the operands of the expression.

Let $D$ be a region of the control flow graph, i.e.\ $D\subseteq BB$ where $BB$ is the set of basic blocks in the control flow graph.

The region $D$ is \textit{valid} if and only if no basic block body in $D$ contains an assignment to any of the operands $v_1,\,v_2,\,\ldots$ of $e$.

The parameters of the optimization are the conditional expression $e$ and any valid set $D$. The transformation is performed in three steps, where the first step is generic code duplication which does not yet consider the conditional expression, the second step rewires some edges to make the other copies reachable and the last step removes the redundant conditionals. Each step preserves the meaning of the program.

\begin{enumerate}
\item (Code duplication) For every basic block $bb_i\in D$, create three copies\footnote{Technically, this is triplication, not duplication.}: the \textit{true copy} $bb_i^t$, the \textit{false copy} $bb_i^f$ and the \textit{unknown copy} $bb_i^u$. The edges of the graph are modified as follows:
\begin{itemize}
\item An edge between $bb_i\notin D$ and $bb_j\notin D$ is left unchanged.
\item An edge between $bb_i\in D$ and $bb_j\in D$ is reproduced by the three edges $bb_i^t$ to $bb_j^t$, $bb_i^f$ to $bb_j^f$ and $bb_i^u$ to $bb_j^u$.
\item An edge between $bb_i\in D$ and $bb_j\notin D$ is reproduced by the three edges $bb_i^t$ to $bb_j$, $bb_i^f$ to $bb_j$ to $bb_i^u$ to $bb_j$.
\item An edge between $bb_i\notin D$ and $bb_j\in D$ is changed to an edge from $bb_i$ to $bb_j^u$.
\end{itemize}
\item (Conditional evaluation) For every conditional edge from $bb_i$ to $bb_j$ depending on $e$ being true (false) at the end of $bb_i$, where $bb_j$ is a copy of a node in $D$, replace it by an edge $bb_i$ to $bb_j^t$ ($bb_j^f$).
\item (Conditional elimination) For every basic block $bb_i\in D$ which has a conditional branch depending on $e$ being true (false), remove the condition in $bb_i^t$ ($bb_i^f$), unconditionally follow the true (false) case and remove the other edge.
\end{enumerate}

This algorithm is correct and safe. For correctness, consider an execution path. If the path does not pass any node in $D$, it is not altered by the above algorithm. If the path passes through $D$, but only through unknown copies, it is also not altered. If the path eventually reaches a true (false) copy of a node, it must be because of an edge altered in step 2. At that point of execution, the value of $e$ is known to be true (false), and because $D$ is valid, it remains so until the execution path leaves the region $D$. Any conditional jump skipped because of step 3 is therefore behaving exactly as in the original execution path.
.

Safeness follows from the fact that we only copy nodes and remove the evaluation of conditionals, so along no path new instructions are added.

\subsection*{Example}

\begin{wrapfigure}{R}{0.4\linewidth}
\begin{algorithm}[H]
\eIf{\ldots}{
\lIf{$e$}{\ldots}\lElse{\ldots}
}{$e \coloneqq \cdots$}
\While{\ldots}{
\lIf{$e$}{\ldots}\lElse{\ldots}
}
{\ldots}
\end{algorithm}
\caption{Example code}
\label{fig:excode}
\end{wrapfigure}

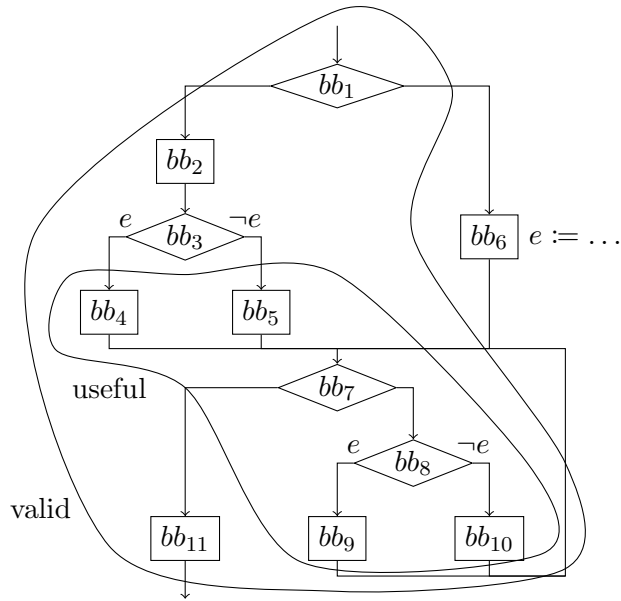
\begin{figure}
\begin{center}
\begin{tikzpicture}
\path[use as bounding box] (-4,0.5) rectangle (4,-6.5);
\path
(0,0) node[draw,shape=diamond,shape aspect=3,inner sep=1pt] (1) {$bb_1$} 
(-2,-1) node[draw,shape=rectangle] (2) {$bb_2$}
(-2,-2) node[draw,shape=diamond,shape aspect=2.5,inner sep=1pt] (3) {$bb_3$}
(-3,-3) node[draw,shape=rectangle] (4) {$bb_4$}
(-1,-3) node[draw,shape=rectangle] (5)  {$bb_5$}
(2,-2) node[draw,shape=rectangle] (6) {$bb_6$} 
(0,-4) node[draw,shape=diamond,shape aspect=2.5,inner sep=1pt] (7) {$bb_7$}
(1,-5) node[draw,shape=diamond,shape aspect=2.5,inner sep=1pt] (8) {$bb_8$}
(0,-6) node[draw,shape=rectangle] (9) {$bb_9$}
(2,-6) node[draw,shape=rectangle] (10) {$bb_{10}$}
(-2,-6) node[draw,shape=rectangle] (11) {$bb_{11}$}
;
\path (6.east) node[right] {$e \coloneqq \ldots$};

\draw[->] (1.north) +(0,.5) -- (1.north) ;
\draw[->] (1.west) -| (2.north);
\draw[->] (1.east) -| (6.north);
\draw[->] (2.south) -| (3.north);
\draw[->] (3.west) -| (4.north) node[above,pos=0] {$e$};
\draw[->] (3.east) -| (5.north) node[above,pos=0] {$\neg e$};
\draw (6.south) |- ($(7.north) +(0,.2)$);
\draw (4.south) |- ($(7.north) +(0,.2)$);
\draw (5.south) |- ($(7.north) +(0,.2)$);
\draw[->]($(7.north) +(0,.2)$) -- (7.north);
\draw[->] (7.west) -| (11.north);
\draw[->] (7.east) -| (8.north);
\draw[->] (8.west) -| (9.north) node[above,pos=0] {$e$};
\draw[->] (8.east) -| (10.north) node[above,pos=0] {$\neg e$};
\draw (10.south) |- ($(10.south) + (1,-0.2)$)
		     |- ($(7.north) +(0,.2)$);
\draw (9.south) |- ($(10.south) + (1,-0.2)$);
\draw[->] (11.south) -- +(0,-.5);

\draw (-3.9,-5.6) node {valid};
\draw plot[smooth cycle] coordinates {(0,1) (1.5,0) (1,-2) (3,-5) (3,-6.4) (0,-6.7) (-3,-6.1) (-4,-2)};

\draw (-3,-4) node {useful};
\draw plot[smooth cycle,dotted] coordinates {(-2,-2.5) (0,-2.5) (2.5,-5) (2.7,-6.2) (-.5,-6.3) (-2,-4) (-3.7,-3.5) (-3.5,-2.5)};

\end{tikzpicture}
\end{center}
\caption{Example control flow graph before CECD}
\label{fig:ex1}
\end{figure}

\begin{figure}
\begin{center}
\begin{tikzpicture}
\begin{scope}[xshift=0cm]
\path
(0,0) node[draw,shape=diamond,shape aspect=3,inner sep=1pt] (1) {$bb_1$} 
(3,-1) node[draw,shape=rectangle] (6) {$bb_6$} 
(-2,-1) node[draw,shape=rectangle] (2) {$bb_2$}
(-2,-2) node[draw,shape=diamond,shape aspect=2.5,inner sep=1pt] (3) {$bb_3$}
(6.west) node[left] {$e \coloneqq \ldots$};
\end{scope}

\path (-5,-7) node[draw,shape=rectangle] (11) {$bb_{11}$};

\begin{scope}[xshift=-3cm]
\path
(-1,-3) node[draw,shape=rectangle] (4t) {$bb_4^t$}
(0,-4) node[draw,shape=diamond,shape aspect=2.5,inner sep=1pt] (7t) {$bb_7^t$}
(1,-5.5) node[draw,shape=rectangle] (9t) {$bb_9^t$}
;
\end{scope}

\begin{scope}[xshift=0cm]
\path
(-1,-3) node[draw,shape=rectangle] (5f)  {$bb_5^f$}
(0,-4) node[draw,shape=diamond,shape aspect=2.5,inner sep=1pt] (7f) {$bb_7^f$}
(1,-5.5) node[draw,shape=rectangle] (10f) {$bb_{10}^f$}
;
\end{scope}

\begin{scope}[xshift=3cm]
\path
(0,-2) node[draw,shape=diamond,shape aspect=2.5,inner sep=1pt] (7u) {$bb_7^u$}
(1,-3) node[draw,shape=diamond,shape aspect=2.5,inner sep=1pt] (8u) {$bb_8^u$}
;
\end{scope}

\draw[->] (1.north) +(0,.5) -- (1.north) ;
\draw[->] (1.west) -| (2.north);
\draw[->] (1.east) -| (6.north);
\draw[->] (2.south) -| (3.north);
\draw[->] (3.west) -| (4t.north) node[above,pos=0] {$e$};
\draw[->] (3.east) -| (5f.north) node[above,pos=0] {$\neg e$};
\draw (6.south) |- ($(7u.north) +(0,.2)$);

\draw (4t.south) |- ($(7t.north) +(0,.2)$);
\draw[->]($(7t.north) +(0,.2)$) -- (7t.north);
\draw[->] (7t.east) -| (9t.north);
\draw (9t.south) |- ($(9t.south) + (-2.2,-0.2)$)
		     |- ($(7t.north) +(0,.2)$);
\draw[->] (7t.west) -| (11.north);

\draw (5f.south) |- ($(7f.north) +(0,.2)$);
\draw[->]($(7f.north) +(0,.2)$) -- (7f.north);
\draw[->] (7f.east) -| (10f.north);
\draw (10f.south) |- ($(10f.south) + (-2.2,-0.2)$)
		     |- ($(7f.north) +(0,.2)$);
\draw (7f.west) |- ($(11.north)+(0,.2)$);

\draw[->]($(7u.north) +(0,.2)$) -- (7u.north);
\draw[->] (7u.east) -| (8u.north);
\draw (8u.west) |- ($(9t.north)+(0,1)$)  node[above,pos=0] {$e$};
\draw (8u.east) |- ($(10f.north)+(0,.2)$)  node[above,pos=0] {$\neg e$};
\draw (7u.west) |- ($(11.north)+(0,.2)$);

\draw[->] (11.south) -- +(0,-.5);
\end{tikzpicture}
\end{center}
\caption{Example control flow graph after conditional evaluation and elimination and dead code elimination}
\label{fig:ex4}
\end{figure}
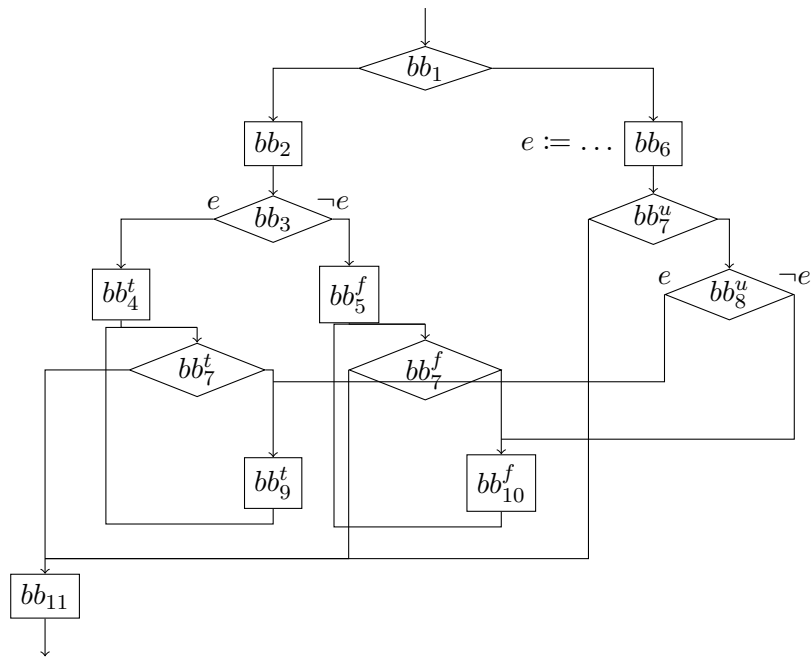

Consider the code fragment in Figure \vref{fig:excode} (leaving out any unrelated assignments or expressions).
The corresponding control flow graph is given in figure \vref{fig:ex1}. The largest valid region is marked, as well as the largest region if useful nodes. Applying the algorithm with $D$ set to the region of useful nodes, after step 1 we obtain the graph shown in figure \vref{fig:ex2}. At this point, the true and false copies are not reachable yet. Steps 2 and 3 modify the edges related to conditional on $e$, and we reach figure \vref{fig:ex3}. This contains a lot of dead code. Removing this in a standard dead code removal pass, we reach the final state \vref{fig:ex4}. It can clearly be seen that on every path from entry to exit, the conditional $e$ is evaluated at most once. Also the issue of a while-loop occurrence (in contrast to the optimizer-friendly do-while-loop) is gracefully taken care of.

\begin{figure}
\begin{center}
\begin{tikzpicture}
\begin{scope}[xshift=4cm]
\path
(0,0) node[draw,shape=diamond,shape aspect=3,inner sep=1pt] (1) {$bb_1$} 
(2,-2) node[draw,shape=rectangle] (6) {$bb_6$} 
(-2,-1) node[draw,shape=rectangle] (2) {$bb_2$}
(-2,-2) node[draw,shape=diamond,shape aspect=2.5,inner sep=1pt] (3) {$bb_3$}
(6.west) node[left] {$e \coloneqq \ldots$};
\end{scope}

\path (-5,-7.5) node[draw,shape=rectangle] (11) {$bb_{11}$};

\begin{scope}[xshift=-4cm]
\path
(-3,-3) node[draw,shape=rectangle] (4t) {$bb_4^t$}
(-1,-3) node[draw,shape=rectangle] (5t)  {$bb_5^t$}
(0,-4) node[draw,shape=diamond,shape aspect=2.5,inner sep=1pt] (7t) {$bb_7^t$}
(1,-5) node[draw,shape=diamond,shape aspect=2.5,inner sep=1pt] (8t) {$bb_8^t$}
(0,-6) node[draw,shape=rectangle] (9t) {$bb_9^t$}
(2,-6) node[draw,shape=rectangle] (10t) {$bb_{10}^t$}
;
\end{scope}

\begin{scope}[xshift=0cm]
\path
(-3,-3) node[draw,shape=rectangle] (4f) {$bb_4^f$}
(-1,-3) node[draw,shape=rectangle] (5f)  {$bb_5^f$}
(0,-4) node[draw,shape=diamond,shape aspect=2.5,inner sep=1pt] (7f) {$bb_7^f$}
(1,-5) node[draw,shape=diamond,shape aspect=2.5,inner sep=1pt] (8f) {$bb_8^f$}
(0,-6) node[draw,shape=rectangle] (9f) {$bb_9^f$}
(2,-6) node[draw,shape=rectangle] (10f) {$bb_{10}^f$}
;
\end{scope}

\begin{scope}[xshift=4cm]
\path
(-3,-3) node[draw,shape=rectangle] (4u) {$bb_4^u$}
(-1,-3) node[draw,shape=rectangle] (5u)  {$bb_5^u$}
(0,-4) node[draw,shape=diamond,shape aspect=2.5,inner sep=1pt] (7u) {$bb_7^u$}
(1,-5) node[draw,shape=diamond,shape aspect=2.5,inner sep=1pt] (8u) {$bb_8^u$}
(0,-6) node[draw,shape=rectangle] (9u) {$bb_9^u$}
(2,-6) node[draw,shape=rectangle] (10u) {$bb_{10}^u$}
;
\end{scope}

\draw[->] (1.north) +(0,.5) -- (1.north) ;
\draw[->] (1.west) -| (2.north);
\draw[->] (1.east) -| (6.north);
\draw[->] (2.south) -| (3.north);
\draw[->] (3.west) -| (4u.north) node[above,pos=0] {$e$};
\draw[->] (3.east) -| (5u.north) node[above,pos=0] {$\neg e$};
\draw (6.south) |- ($(7u.north) +(0,.2)$);

\draw (4t.south) |- ($(7t.north) +(0,.2)$);
\draw (5t.south) |- ($(7t.north) +(0,.2)$);
\draw[->]($(7t.north) +(0,.2)$) -- (7t.north);
\draw[->] (7t.east) -| (8t.north);
\draw[->] (8t.west) -| (9t.north) node[above,pos=0] {$e$};
\draw[->] (8t.east) -| (10t.north) node[above,pos=0] {$\neg e$};
\draw (10t.south) |- ($(9t.south) + (0,-0.2)$);
\draw (9t.south) |- ($(9t.south) + (-1.2,-0.2)$)
		     |- ($(7t.north) +(0,.2)$);
\draw[->] (7t.west) -| (11.north);

\draw (4f.south) |- ($(7f.north) +(0,.2)$);
\draw (5f.south) |- ($(7f.north) +(0,.2)$);
\draw[->]($(7f.north) +(0,.2)$) -- (7f.north);
\draw[->] (7f.east) -| (8f.north);
\draw[->] (8f.west) -| (9f.north) node[above,pos=0] {$e$};
\draw[->] (8f.east) -| (10f.north) node[above,pos=0] {$\neg e$};
\draw (10f.south) |- ($(9f.south) + (0,-0.2)$);
\draw (9f.south) |- ($(9f.south) + (-1.2,-0.2)$)
		     |- ($(7f.north) +(0,.2)$);
\draw (7f.west) |- ($(11.north)+(0,.2)$);

\draw (4u.south) |- ($(7u.north) +(0,.2)$);
\draw (5u.south) |- ($(7u.north) +(0,.2)$);
\draw[->]($(7u.north) +(0,.2)$) -- (7u.north);
\draw[->] (7u.east) -| (8u.north);
\draw[->] (8u.west) -| (9u.north) node[above,pos=0] {$e$};
\draw[->] (8u.east) -| (10u.north) node[above,pos=0] {$\neg e$};
\draw (10u.south) |- ($(9u.south) + (0,-0.2)$);
\draw (9u.south) |- ($(9u.south) + (-1.2,-0.2)$)
		     |- ($(7u.north) +(0,.2)$);
\draw (7u.west) |- ($(11.north)+(0,.2)$);

\draw[->] (11.south) -- +(0,-.5);
\end{tikzpicture}
\end{center}
\caption{Example control flow graph after code duplication of useful nodes}
\label{fig:ex2}
\end{figure}
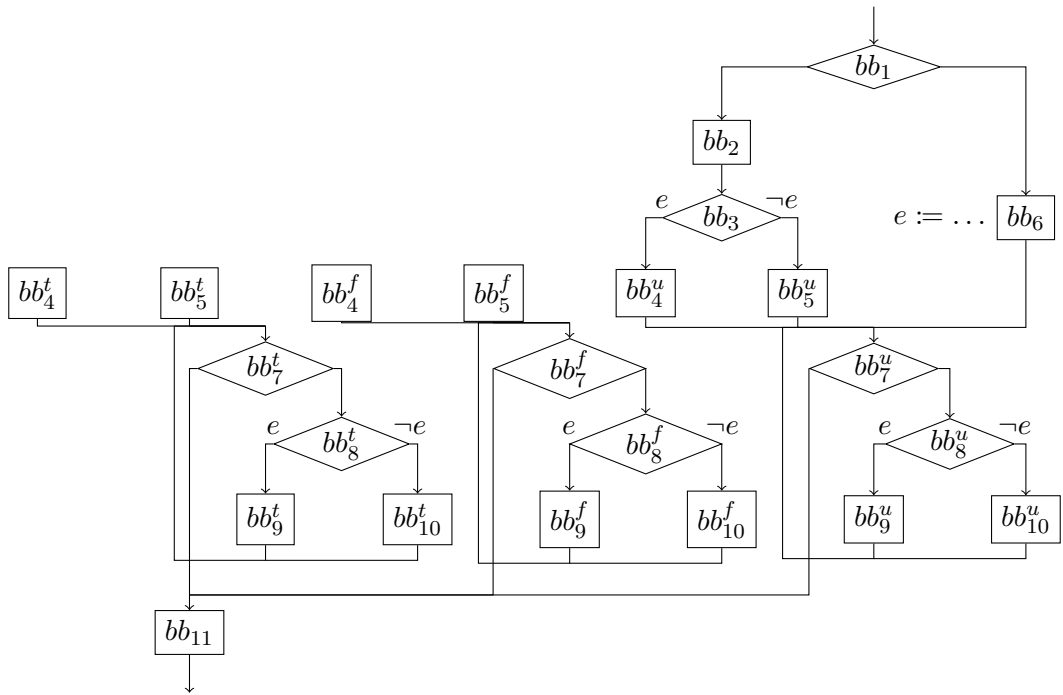

\begin{figure}
\begin{center}
\begin{tikzpicture}
\begin{scope}[xshift=4cm]
\path
(0,-0.5) node[draw,shape=diamond,shape aspect=3,inner sep=1pt] (1) {$bb_1$} 
(2,-2) node[draw,shape=rectangle] (6) {$bb_6$} 
(-8,-1) node[draw,shape=rectangle] (2) {$bb_2$}
(-8,-2) node[draw,shape=diamond,shape aspect=2.5,inner sep=1pt] (3) {$bb_3$}
(6.west) node[left] {$e \coloneqq \ldots$};
\end{scope}

\path (-5,-7.5) node[draw,shape=rectangle] (11) {$bb_{11}$};

\begin{scope}[xshift=-4cm]
\path
(-3,-3) node[draw,shape=rectangle] (4t) {$bb_4^t$}
(-1,-3) node[draw,shape=rectangle] (5t)  {$bb_5^t$}
(0,-4) node[draw,shape=diamond,shape aspect=2.5,inner sep=1pt] (7t) {$bb_7^t$}
(1,-6) node[draw,shape=rectangle] (9t) {$bb_9^t$}
;
\end{scope}

\begin{scope}[xshift=0cm]
\path
(-3,-3) node[draw,shape=rectangle] (4f) {$bb_4^f$}
(-1,-3) node[draw,shape=rectangle] (5f)  {$bb_5^f$}
(0,-4) node[draw,shape=diamond,shape aspect=2.5,inner sep=1pt] (7f) {$bb_7^f$}
(1,-6) node[draw,shape=rectangle] (10f) {$bb_{10}^f$}
;
\end{scope}

\begin{scope}[xshift=4cm]
\path
(-3,-3) node[draw,shape=rectangle] (4u) {$bb_4^u$}
(-1,-3) node[draw,shape=rectangle] (5u)  {$bb_5^u$}
(0,-4) node[draw,shape=diamond,shape aspect=2.5,inner sep=1pt] (7u) {$bb_7^u$}
(1,-5) node[draw,shape=diamond,shape aspect=2.5,inner sep=1pt] (8u) {$bb_8^u$}
(0,-6) node[draw,shape=rectangle] (9u) {$bb_9^u$}
(2,-6) node[draw,shape=rectangle] (10u) {$bb_{10}^u$}
;
\end{scope}

\draw[->] (1.north) +(0,.5) -- (1.north) ;
\draw[->] (1.west) -| (2.north);
\draw[->] (1.east) -| (6.north);
\draw[->] (2.south) -| (3.north);
\draw[->] (3.west) -| (4t.north) node[above,pos=0] {$e$};
\draw[->] (3.east) -| (5f.north) node[above,pos=0] {$\neg e$};
\draw (6.south) |- ($(7u.north) +(0,.2)$);

\draw (4t.south) |- ($(7t.north) +(0,.2)$);
\draw (5t.south) |- ($(7t.north) +(0,.2)$);
\draw[->]($(7t.north) +(0,.2)$) -- (7t.north);
\draw[->] (7t.east) -| (9t.north);
\draw (9t.south) |- ($(9t.south) + (-2.2,-0.2)$)
		     |- ($(7t.north) +(0,.2)$);
\draw[->] (7t.west) -| (11.north);

\draw (4f.south) |- ($(7f.north) +(0,.2)$);
\draw (5f.south) |- ($(7f.north) +(0,.2)$);
\draw[->]($(7f.north) +(0,.2)$) -- (7f.north);
\draw[->] (7f.east) -| (10f.north);
\draw (10f.south) |- ($(10f.south) + (-2.2,-0.2)$)
		     |- ($(7f.north) +(0,.2)$);
\draw (7f.west) |- ($(11.north)+(0,.2)$);

\draw (4u.south) |- ($(7u.north) +(0,.2)$);
\draw (5u.south) |- ($(7u.north) +(0,.2)$);
\draw[->]($(7u.north) +(0,.2)$) -- (7u.north);
\draw[->] (7u.east) -| (8u.north);
\draw[->] (8u.west) -| (9t.north) node[above,pos=0] {$e$};
\draw[->] (8u.east) -| ($(10u.north)+(0,0.2)$) node[above,pos=0] {$\neg e$} -| (10f.north) ;
\draw (10u.south) |- ($(9u.south) + (0,-0.2)$);
\draw (9u.south) |- ($(9u.south) + (-1.2,-0.2)$)
		     |- ($(7u.north) +(0,.2)$);
\draw (7u.west) |- ($(11.north)+(0,.2)$);

\draw[->] (11.south) -- +(0,-.5);
\end{tikzpicture}
\end{center}
\caption{Example control flow graph after conditional evaluation and elimination}
\label{fig:ex3}
\end{figure}
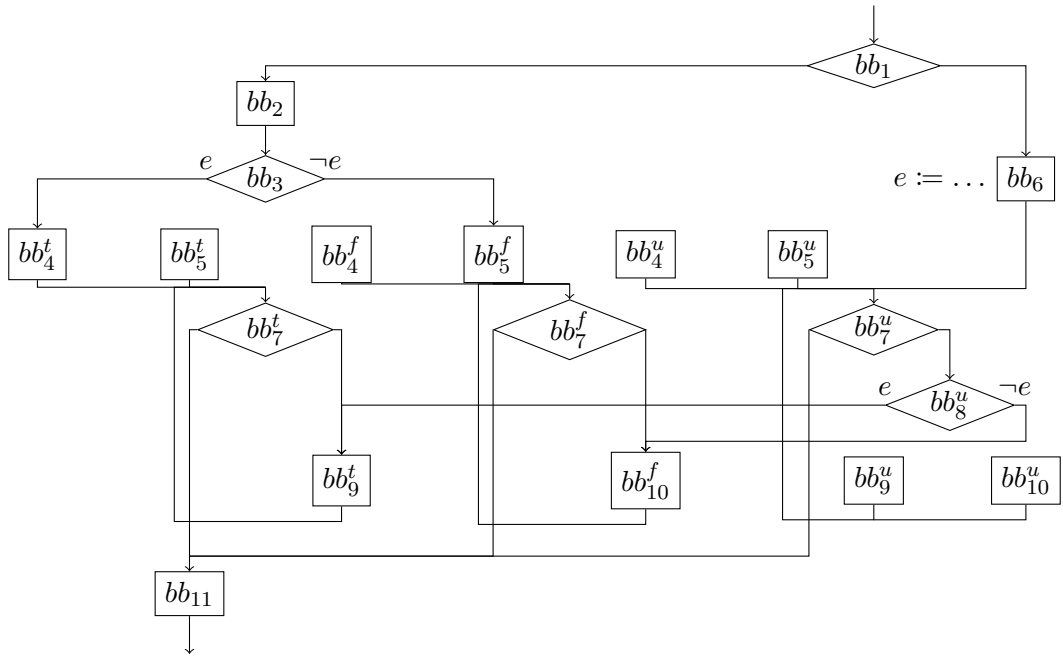

\section{The region of duplication}
\label{sec:region}

The above algorithm works for any valid region, and validity is a simple local property that is easily checked. But not all valid regions are useful. For example, entry nodes $bb_i$ of the region where no incoming edge depends on $e$ would be duplicated, but only $bb_i^u$ would be reachable. Similarly, exit nodes of the region that do not have a conditional evaluation of $e$ would be copied for no gain.

\subsection{Usefulness}

Therefore, we can define that a node $bb_i$ in a valid region $D$ to be \textit{useless} if
\begin{itemize}
\item on all paths leading to $bb_i$, there is no conditional evaluation of $e$ followed only by nodes in $D$ or
\item no path originating from $bb_i$ reaches an conditional evaluation of $e$ before it leaves the region $D$.
\end{itemize}

A node $bb_i\in D$ that is not useless is \textit{useful}.

Uselessness is, in contrast to validity, not a property of the basic block alone but defined with respect to the chosen region $D$. A basic block may be useless in $D$ but not so in a different region $D'$. But the property is monotonous: If $D'\subseteq D$ and $D$ is useful in $D'$, then it is also useful in $D$.

\subsection{Evaluation of a region}
\label{sec:heurisitc}

For a given conditional expression, there are many possible regions of duplication, and even if we only consider fully useful regions, their number might be exponential in the size of the graph. Therefore we need an heuristic that selects a sensible region or decides that no region is good enough to perform CECD. We split this decision into two independent steps: Region Selection, where the the best region for a particular conditional, for some meaning of “best” is chosen, and Region Evaluation, where it is decided whether CECD should be performed for the selected region.

These decisions have to depend on the intended use of the code. Code for an embedded system might have very tight size requirements and large regions of duplication would be unsuitable, whereas code written for massive numerical calculations may be allowed to grow quite a bit if it removes instructions from the inner loops.

At this point, we suggest a very simple heuristic for Region Selection: To cover as many executions paths as possible, we just pick the largest valid region consisting of useful nodes. The heuristic for Region Evaluation expects one parameter $k$, which is the number of additional expressions that the program is allowed to grow for one conditional to be removed. Together, this amounts to the following steps being taken:

\begin{enumerate}
\item Let $D$ be the largest valid region consisting only of useful nodes.
\item Let $R^t$, $R^f$ resp.\ $R^u$ the set of those basic blocks in $D$, whose true, false resp\. unknown copy will be reachable after CECD.
\item Let $n$ be the number of basic blocks in $D$ that contain a conditional evaluation of $e$, i.e.\ the number of redundant conditionals.
\item If
\[
\sum_{bb_i\in R^t} S(bb_i) + 
\sum_{bb_i\in R^f} S(bb_i) + 
\sum_{bb_i\in R^u} S(bb_i) - 
\sum_{bb_i\in D} S(bb_i) \le n \cdot k,
\]
where $k$ is a user-defined parameter and $S(bb_i)$ is the number of instructions in the basic block $bb_i$, perform CECD on $D$, otherwise do not perform CECD for this conditional expression.
\end{enumerate}

A number of improvements to this scheme come to mind:
\begin{itemize}
\item The selection heuristic should consider subsets of the largest valid and useful regions as well.
\item It should give different weights to conditionals that are completely removed and conditionals that are only partially removed.
\item Removal of conditionals in inner loops should allow for a larger increase of code size.
\item Given sufficiently detailed execution traces, a more exact heuristic can be implemented. In the next section we see that this easily leads to a $\mathcal{NP}$-hard problem.
\end{itemize}

\subsection{$\mathcal{NP}$-hardness of a profiling based Region Selection heuristic}

A straight forward extension of the above Region Selection heuristic that takes profiling data in the form of execution traces into account, would maximize the sum $\sum_{bb_i\in E} f(bb_i)$, where $E$ is the set of of basic blocks containing an eliminated conditional and $f(bb_i)$ is the number of paths in the execution traces where the conditional in $bb_i$ would be eliminated due to CECD. For simplicity, we assume that an occurrence of a conditional expression does not contribute to the size $S(bb_i)$ of a basic block.

If we have an algorithm that selects the optimal region, we can solve the 0-1 knapsack problem, which is $\mathcal{NP}$-complete. The specification of this problem is as follows:
\begin{quote}
Given $n$ items with weight $w_i\in \mathbb N$ and value $v_i\in \mathbb N$, $i=1,\ldots,n$ and a bound $W\in\mathbb N$, find a selection of items $X \subseteq \{1,\ldots,n\}$ that maximizes the sum $\sum_{i\in X} v_i$ under the constraint $\sum_{i\in X} w_i \le W$.
\end{quote}
Given such a problem, we construct a control flow graph and profiling data as follows:
\begin{itemize}
\item The entry node is $bb_s$, which contains a conditional expression $e$. Both conditional branches point to the node $bb_r$.
\item There is one exit node $bb_e$ with a conditional expression $e$.
\item The node $bb_r$ is the root of a binary tree of basic blocks. The inner nodes contain no instructions but conditional jumps with conditional expressions that are pairwise distinct and distinct from $e$.
\item The tree contains $n$ leaf nodes $bb_l^i$, $i=1,\ldots,n$. The node $bb_l^i$ contains $w_i$ instructions, i.e.\ $S(bb_l^i)=w_i$ and the profiling data gives a frequency of $v_i$ for the execution path passing through $bb_l^i$.
\item The parameter $k$ is chosen to be $W$.
\end{itemize}

A valid and useful region of duplication $D$ in this CFG corresponds to a subset of $X \in {1,..,n}$ and, if non-empty, includes $bb_e$, $bb_l^i$ for $i\in X$ and the nodes connecting $bb_r$ with those leaf nodes. Because $bb_s$ dominates all nodes in $D$, no unknown copies will be generated, and both true and false copies are reachable. The inner nodes of the binary tree and $bb_e$ only contain conditional expressions and thus do not contribute to the size of the duplicated region. Only one redundant conditional occurs, hence $n=1$. The number of executions of $bb_e$ where the conditional is eliminated is exactly the number of execution paths that pass through one of the leaf nodes in $D$. Therefore, the constraint imposed by the Region Evaluation heuristic becomes
\begin{align*}
\sum_{bb_i\in R^t} S(bb_i) + 
\sum_{bb_i\in R^f} S(bb_i) + 
\sum_{bb_i\in R^u} S(bb_i) - 
\sum_{bb_i\in D} S(bb_i) &\le n \cdot k \tag*{$\iff$} \\
\sum_{i\in X} S(bb_l^i) + \sum_{i\in X} S(bb_l^i) + 0 - \sum_{i \in X} S(bb_l^i) &\le 1 \cdot k  \tag*{$\iff$} \\
\sum_{i\in X} w_i &\le W
\end{align*}
and the term to be optimized can be transformed as follows:
\begin{align*}
\sum_{bb_i\in E} f(bb_i) 
= \sum_{i\in X} f(bb_l^i) 
= \sum_{i\in X} v_i.
\end{align*}

This concludes the proof of $\mathcal{NP}$-hardness of this profiling-based heuristic for CECD.

The assumption that conditional expressions do not contribute to the size of a node is not critical: If they do contribute, then this result can still be obtained by a technical modification: Increase $k$ by one and then scale $k$ and the number of instructions in the nodes $bb_l^i$ by a factor larger than the number of all conditional expressions occurring.

\section{Data Flow equations}
\label{sec:df}

\newcommand{\V}{\text{Valid}}
\newcommand{\T}{\text{TrueEdge}}
\newcommand{\F}{\text{FalseEdge}}
\newcommand{\E}{\text{Expr}}
\renewcommand{\L}{\text{Live}}
\newcommand{\A}{\text{Antic}}
\renewcommand{\succ}[1]{\operatorname{succ}(#1)}
\newcommand{\pred}[1]{\operatorname{pred}(#1)}

Three properties of basic blocks have been defined so far: Validness, usefulness and, for the heuristics, which copies of the block will be present after dead code removal. The first one is a purely local property, while the others can be obtained by standard data flow analyses. The defining equations are given in this section. $\succ{i}$ is the set of successor nodes of $bb_i$ in the control flow graph, $\pred{i}$ the set of predecessors. We assume that nodes with a conditional jump have exactly two successors, one for true and one for false.

Local properties:
\begin{itemize}
\item $\V_i$: Basic block $bb_i$ does not contain an assignment to an operator of $e$.
\item $\T_{ij}$: An edge $bb_i \to bb_j$ exists and depends on $e$ being true.
\item $\F_{ij}$: An edge $bb_i \to bb_j$ exists and depends on $e$ being false.
\item $\E_{i} = \sum_{j\in\succ{i}} \T_{ij} + \F_{ij}$: $e$ is a conditional expression in $bb_i$
\end{itemize}

Determining the largest valid region $D$ of useful nodes:
\begin{itemize}
\item $\L_i = \V_i \cdot \sum_{j\in \pred{i}} \E_j + \L_{j}$
\item $\A_i = \V_i \cdot (\E_i + \sum_{j\in \succ{i}} \A_j)$
\item $D_i = \L_i \cdot \A_i$
\end{itemize}

Given a valid region $D$ (which may or may not be obtained using our suggested simple heuristic), determining which copies of the nodes therein are reachable:
\begin{itemize}
\item $R^u_i = D_i \cdot \sum_{j\in \pred{i}} \neg \E_j\cdot(\neg D_j + R^u_j)$
\item $R^t_i = D_i \cdot \sum_{j\in \pred{i}} R^t_j + \T_{ji}$
\item $R^f_i = D_i \cdot \sum_{j\in \pred{i}} R^f_j + \F_{ji}$
\end{itemize}

All given data flow equations are any-path equations and therefore, the values can be initialized to \textit{false} before solving the equations using a standard iterative round-robin or worklist approach.

\section{Future work and conclusions}

While the “how” of CECD is fully understood, the question of “where” and “when”, i.e.\ coming up with good heuristics for the selection of the conditional and region of duplication, needs much further investigation. Also, experiments with real code have yet to be conducted to quantify the benefit and suggest good values for the heuristics’ parameters. Another possible improvement would be to not only consider syntactically equal conditions, but also take algebraic identities into account.

The simplicity of the CECD transformation and the fact that it can easily handle complex control flow indicate that it could be an optimization of general interest.

\bibliographystyle{alphadin}
\bibliography{CECD}

\end{document}